**Aryl Functionalization as a Route to Band Gap Engineering in Single Layer Graphene Devices**

Hang Zhang[1], Elena Bekyarova[2], Jhao-Wun Huang[1], Zeng Zhao[1], Wenzhong Bao[1], Fenglin Wang[1], Robert C. Haddon[2,3*] and Chun Ning Lau[1*]

[1]Department of Physics and Astronomy, [2]Department of Chemistry, [3]Department of Chemical and Environmental Engineering, University of California, Riverside, CA 92521

*Emails: haddon@ucr.edu, lau@physics.ucr.edu

**TABLE OF CONTENT GRAPHIC**

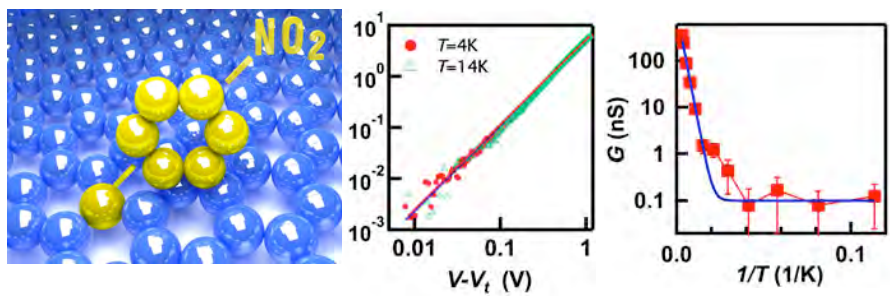

**ABSTRACT**

Chemical functionalization is a promising route to band gap engineering of graphene. We chemically grafted nitrophenyl groups onto exfoliated single-layer graphene sheets in the form of substrate-supported or free-standing films. Our transport measurements demonstrate that non-suspended functionalized graphene behaves as a granular metal, with variable range hopping transport and a mobility gap ~ 0.1 eV at low temperature. For suspended graphene that allows functionalization on both surfaces, we demonstrate tuning of its electronic properties from a granular metal to a gapped semiconductor, in which charge transport occurs via thermal activation over a gap ~ 80 meV. This non-invasive and scalable functionalization technique paves the way for CMOS-compatible band gap engineering of graphene electronic devices.

Since its experimental isolation on insulating substrates[1] and the development of a wafer scale growth technology[2], graphene has rapidly become a promising candidate for next generation electronic material[3]. Its material properties, such as atomically thin dimension, unparalleled room-temperature mobility[4-8], thermal conductivity[9] and current carrying capacity[10], are far superior to those of silicon, whereas its two-dimensionality (2D) is naturally compatible with standard CMOS-based technologies. However, as a gapless semiconductor, graphene cannot be directly applied in standard digital electronic circuitry. Thus, in order to realize its potential to supplement or replace Si, band gap creation and control in graphene still poses a major challenge. Various approaches have been proposed and implemented, such as the use of biased bilayers[11, 12], graphene nanoribbons[13, 14] and strain-based device fabrication[15, 16]. Chemical functionalization[17-20], because of its simple implementation and potential compatibility with wafer-scale heterogeneous integration, is a particularly promising approach. Prior work in this area includes hydrogenation to form graphane[21, 22], a wide band gap semiconductor, and fluorination to form insulators[23-25]. However, such functionalization procedures are invasive, thermally unstable and highly energetic processes that are incompatible with standard semiconductor technology.

Here we report the successful aryl-group functionalization[26] and transport measurements of single layer graphene sheets that are isolated in the form of free-standing films or supported on Si/SiO$_2$ substrates. Functionalization of graphene is verified by the development of $D$ and $D^*$ peaks in Raman spectroscopy measurements[27, 28]. The chemical treatment is simple, non-invasive and leads to the formation of stable $sp^3$ C-C bonds to the basal plane of graphene, and can be easily scaled up for industrial production. The devices' field effect mobility after functionalization is ~ 50 cm$^2$/Vs for substrate-supported graphene and ~200 for the suspended films, comparable with that of Si. By performing electrical measurements at different temperatures, we demonstrate that non-suspended functionalized (NSF) graphene behaves as a granular metal, displaying variable range hopping (VRH) transport, a localization-induced gap ~0.1V at 4K and a charge localization length of 40-125 nm. In the case of the suspended devices which allow double-sided grafting of the aryl groups, we obtain a variable degree of functionalization, and consequently the transport mechanism spans a large range: charges traverse across lightly functionalized devices via VRH, while heavily functionalized devices behave as gapped semiconductors, in which charges are thermally activated over a gap of ~ 80meV at high temperature, crossing over to quantum tunneling for $T<30K$. Our results demonstrate the potential for tunable electronic properties of graphene via non-invasive solution chemistry, paving the way for wafer-scale manipulation of band gap and electronic properties of graphene.

Graphene sheets were extracted from bulk graphite using standard mechanical exfoliation techniques, and coupled to Cr/Au electrodes via electron beam lithography. The devices are then annealed in vacuum by applying a large current[29] or using a local heater. A scanning electron microscope (SEM) image of a typical non-suspended device is shown in the Figure 1a inset. Immediately after fabrication, these pristine devices are electrically characterized at different temperatures. Figure 1a displays the current-voltage (*I-V*) characteristics of one such device NS1, which is nearly linear up to 1V. With the modulation of the gate voltage $V_g$, which changes the charge density induced in the graphene sheet, the devices' four terminal differential conductance *G* at zero bias increases approximately linearly away from the Dirac point, which is at $V_D$=9V, indicating a field effect mobility of $\mu$~2000 cm$^2$/Vs (Figure 1b). Both the *I(V)* and *G(V_g)* curves display minimal temperature dependence, in agreement with previous experiments[4-6].

These devices are then chemically modified by immersing in an acetonitrile solution of 4-nitrophenyl diazonium tetrafluoroborate ($NO_2$-$C_6H_4N_2^+BF_4^-$, 10 mM) in the presence of tetrabutylammonium hexafluorophosphate ($[Bu_4N]PF_6$, 0.1 M). The spontaneous grafting (Figure 1c) was performed in a glove box in the absence of light; the devices were then rinsed with acetonitrile, acetone and isopropanol and dried in air. The functionalization process was confirmed by Raman spectroscopy as discussed in previous publications[28].

Figure 1d shows the evolution of the Raman spectra before and after chemical attachment of aryl groups to the carbon atoms of the graphene lattice. Pristine graphene exhibits two prominent Raman peaks – a *G*-band at ~ 1580 cm$^{-1}$ associated with the doubly degenerate phonon mode ($E_{2g}$) at the Brillouin zone center, and the 2*D*-band at ~2700 cm$^{-1}$ that originates from a second order double resonance process. The intensity of the 2*D*-band relative to the *G*-band is characteristic of monolayer graphene. In some pristine graphene samples, the *G*-band was observed at 1590 cm$^{-1}$; such blue-shift of *G*-peak suggests *p*-doping, which is presumably due to environmental oxygen, is confirmed by transport measurements. After chemical treatment, the Raman spectra of all of the devices developed strong *D*-bands at ~1340 cm$^{-1}$, accompanied by small *D\**-bands at 1620 cm$^{-1}$ and *D+D\** bands; the reduction in intensity of the 2D band is indicative of doping of the films by the reagent[30]. The functionalized devices are stable up to more than 200 ºC.

After functionalization, the devices are annealed again at $T<\approx 150K$. Transport characteristics through the NSF devices are dramatically altered from those of pristine graphene. As shown by the red traces in Figure 2a-b, at 300K, the *I-V* curves of device NS1 are approximately linear; however, the device conductance decreases by a factor of 20 from that of pristine graphene, while $\mu$ decreases to ~ 50 cm$^2$/Vs. At low temperature *T*=4.2K, the *I-V* characteristics becomes markedly non-linear. For small bias, the conductance is effectively zero for all values of $V_g$; at higher bias, the $G(V_g)$ curve recovers its high temperature behavior, albeit with ~20% reduction in conductance. We note that at 4K, the Dirac point shifts to 38V, presumably due to random charge fluctuations in the electromagnetic environment of the device.

At low temperature, the most notable feature of the device's behavior is the zero or extremely small conductance at small bias. Figure 2c displays the differential conductance $dI/dV$ of the device (color), which is obtained by numerically differentiating the $I$-$V$ curves taken at different $V_g$ value, as functions of $V_g$ (horizontal axis) and $V$ (vertical axis). Two line traces, $dI/dV$ vs. $V$ at the Dirac point $V_g$=38V and at the highly hole-doped regime $V_g$=-28.5V, are shown in Figure 2d. The blue region, which has effectively zero conductance (<0.1 µS), appears as a half diamond. Similar diamond-shaped low $G$ regions have been observed in extremely narrow graphene nanoribbons[14], and suggests the formation of a transport gap or mobility gap at low temperature. From the height of region with $dI/dV$<0.1 µS in Figure 2c, we estimate the magnitude of the transport gap to be ~0.1 eV.

In order to elucidate the transport mechanism of the device, we investigated the temperature dependence of its $I$-$V$ characteristics. For transport across a true band gap $\Delta$, one expects thermally activated behavior, $G \propto e^{-\Delta/2k_BT}$, where $k_B$ is Boltzmann constant. Figure 3a displays the $I$-$V$ characteristics of device NS1 taken at different temperatures between 14 K and 300K, which exhibits a transition from non-linear to linear $IV$ as $T$ increases. The low bias conductance $G$ in the linear response regime is extracted and plotted in Figure 3b as a function of $1/T$, where the red and green data points corresponding to those taken at the Dirac point and at highly electron-doped regime, respectively. The non-linearity of the data excludes thermal activation as the transport mechanism.

Another possible mechanism is variable range hopping (VRH)[31-33], in which the electrons in a strongly disordered or granular system traverse the system via a series of hops to neighboring localized states. The hopping distance is determined by a competition that maximizes the wave function overlap and minimizes the activation energy between the two sites. As a result, the zero bias conductance has a stretched-exponential dependence on $T$

$$G = A\exp\left[-\left(\frac{T_0}{T}\right)^\alpha\right] \qquad (1)$$

where $A$ and $T_0$ are constants. The exponent $\alpha$, which depends on the dimensionality and anisotropy of the system, strength of disorder and dielectric constant, has been theoretically and experimentally determined to vary from ¼ to 3/4 in systems as varied as carbon nanotube networks[34], polymers[35], arrays of nanoparticles[36, 37] and the 2D electron gas in the quantum Hall regime[38]. Generally, there exist two different VRH regimes. In the Mott VRH regime[32], the system has a constant density of states (DOS) near the Fermi level, and negligible Coulomb interaction between the hopping sites. This regime dominates at high temperature, yielding $\alpha=1/(d+1)$, where $d$ is the dimensionality of the system. At sufficiently low temperatures, Coulomb interaction between the sites gives rise to a Coulomb gap at the Fermi level; in this Efros-Shklovskii (ES) regime, $\alpha=1/2$ for all $d$[31]. Since $d=2$ for graphene, we expect $\alpha=1/3$ in the Mott regime and $\alpha=1/2$ in the ES regime. The ES VRH model also applies for granular metals[39, 40], where transport occurs via multiple electron co-tunneling among different grains[40].

To quantitatively account for our results, we fit the data in Figure 3b to Eq. (1), as shown by the solid lines in Figure 3b. Excellent agreement with data over the entire $T$ range was obtained – for the data set at the Dirac point, the best-fit parameters are $A=86.9$, $T_0=1670$K and $\alpha=0.39$; for the data set at highly doped regime, $A=42.2$, $T_0=317$K and $\alpha=0.42$. For both regimes,

$\alpha$ is determined to be ~ 0.4, which may indicate a crossover from the Mott to ES VRH regimes; alternatively, it could suggest that the DOS $g(\varepsilon) \sim \varepsilon^{1/2}$ for functionalized graphene[33], where $\varepsilon$ is the energy measured relative to the Fermi level. Further experimental and theoretical investigation of functionalized graphene will be necessary to ascertain the exponent and understand the rich underlying physics.

At low temperatures the system is expected to be in the ES regime; as shown in Figure 2c, the data points in the range T =14 – 150K fall on a straight line in the log plot of $G$ against $T^{-1/2}$ ( adequate but slightly more non-linear traces are obtained if we plot against $T^{-1/3}$). The data are well fit by the function $G = A\exp\left(-\sqrt{T_0/T}\right)$, with a $T_0$ value of 450K and 150K in the DP and highly doped regime, respectively. Since $T_0$ is related to electron localization length $\xi$[31],

$$k_B T_0 \approx C \frac{e^2}{4\pi\varepsilon_0 \varepsilon_r \xi}, \qquad (2)$$

we infer $\xi \approx 42$ nm at the DP and in the doped regime, $\xi \approx 125$ nm. Here $e$ is electron charge, $\varepsilon_0$ and $\varepsilon_r \approx 2.5$ are the permittivity of vacuum and of the electromagnetic environment of the device, respectively, and $C \sim 2.8$ is a constant.

To check the self-consistency of the model, we also estimate the size of the Coulomb gap $\Delta_C$ and the onset temperature $T^*$ of the ES regime. In 2D, $\Delta_C \sim \dfrac{e^4 g_0}{(4\pi\varepsilon_0 \varepsilon_r)^2}$,

$k_B T^* \sim \dfrac{e^6}{(4\pi\varepsilon_0 \varepsilon_r)^3} g_0^2 \xi$, where $g_0$ is the "bare" DOS at the Fermi level. We note that these expressions contain unknown prefactors of order unity, which need to be determined by computer simulations. $g_0$ for our devices are not known, but we can obtain an order-of-magnitude estimate by using graphene's DOS at charge density $n\sim 5\times 10^{11}$ cm$^{-2}$, which is the typical doping level induced by impurities. Given $g_0=2\sqrt{\dfrac{n}{\pi}}\dfrac{1}{\hbar v_F}$, where $v_F \sim 10^6$ m/s is the Fermi velocity, we find that $\Delta_C\sim 500$ K, which is consistent with our observation of a transport gap ~100 meV, and $T^*\sim 3000$K. Considering the uncertainties in the prefactors and $g_0$, these values are indeed reasonable, and also consistent with those found in other 2D systems[41, 42].

The agreement between our data and VRH model, together with the relatively small localization length, suggest that at low temperature, aryl-functionalized graphene behaves as a granular metal. Thus, the NSF device can be modeled as a 2D array of metallic islands; theoretical predictions for such a system[43] suggest that at low temperature, disorder and charging effects lead to a conduction threshold $V_t$, below which $G=0$. At large bias $V>V_t$, the charges percolate through the array via multiple branching paths which optimize the total charging energy, giving rise to a current with a power-law dependence on the reduced voltage

$$I \sim (V-V_t)^{\gamma} \qquad (3)$$

where $\gamma$ is predicted to be 5/3 for a 2D array and the threshold voltage $V_t$ is expected to decrease linearly with increasing $T$[43]. For a given system, Eq. (3) is expected to be universal at low temperatures.

In Figure 3d, we plot $I$ vs. $V-V_t$ at $T=4$K and 14K; the data are taken at the DP, and $V_t$ is chosen to be 80mV and 55mV, respectively. On the log-log plot, both curves collapse into a single straight line, indicating a power-law behavior. The best-fit exponent $\gamma$ is found to be 1.66, in excellent agreement with the theoretical value of 1.6. This establishes that aryl-grafted graphene is a granular metal, with a localization induced transport gap ~ 100mV at $T=4$K. Furthermore, since disordered arrays often display $IV$ characteristics that have a variety of $\gamma$ values, and sometimes even deviate from the simple power law behavior[37, 44], the excellent fit of Eq. (3) to the data suggests that the grafting sites are relatively well-ordered.

Finally, we focus on transport in chemically functionalized free-standing graphene membranes. Since both sides of these membranes are exposed to the aryl solution, we expect more effective grafting due to the minimization of incommensurate ordering and relaxation of strains that result from the *sp²*-to-*sp³* conversion. Moreover, functionalization of suspended graphene is critical for chemical and biological sensors based on graphene nanomechanical resonators, which promise to be highly sensitive as well as selective.

To this end, we functionalized suspended graphene devices obtained by acid release of the underlying $SiO_2$ layer [7, 8] (Figure 4a). The mobility of a typical pristine suspended device is ~5,000 to 15,000 cm²/Vs (Figure 4b). After chemical treatment, the device mobility decreases significantly, to ~ 1-200 cm²/Vs. A lightly functionalized suspended graphene sample displays similar VRH transport characteristics to those of the NSF devices. In contrast, transport across a heavily functionalized suspended device is dramatically different. Figure 4c,d display the electrical characteristics of a device after 20 hours of functionalization. Its *I-V* curves are non-linear even at 300K; at 4K, the conductance is effectively zero (< 0.01 nS) for *V*<1V (Figure 4c). Its zero bias conductance decreases exponentially with 1/*T* at high temperature, and crosses over to a constant value for *T*<30K (Figure 4d). The data can be satisfactorily described by the equation $G(T)=G_0+Aexp(-E_A/k_BT)$, where $E_A \approx 39 \pm 10$ meV is the activation energy and $G_0$ is the constant background conductance, which is likely the noise floor of this measurement setup. This activation energy may arise from the formation of a true band gap – in fact, our previous study using ARPES measurement has shown the presence of a band gap of ~ 360 meV[28]; alternatively, the Arrhenious behavior may arise from nearest neighbour hopping. Thus, we demonstrate the formation of a gap $2E_A$ ~80 meV at room temperature in graphene sheets that are functionalized on both surfaces, though further experimental work will be necessary to ascertain the nature and magnitude of the gap.

In conclusion, we have demonstrated single- and double-sided chemical functionalization of graphene sheets via grafting of aryl groups using simple solution chemistry. Via control of the functionalization, graphene can be tuned from a gapless semi-metal to granular metal with a Coulomb-induced transport gap, and to a semiconductor with thermal activation energy ~100 meV. Since the aryl-grafting technique is non-invasive and compatible with CMOS technologies, our results have significant impact for band gap engineering and control in graphene electronics.

**Acknowledgment**

We thank Marc Bockrath for helpful discussions. This work was supported in part by ONR/DMEA H94003-10-2-1003, NSF CAREER DMR/0748910, ONR N00014-09-1-0724, NSF-MRSEC DMR-0820382 and the FENA Focus Center.

FIGURE 1. (a,b). *I-V* and *G(V_g)* characteristics of a pristine graphene device NS1 on Si/SiO$_2$ substrate at 300K (red traces) and 4K (blue traces). Inset in (a): SEM image of a device. Scale bar: 5 μm. (c). Schematics of the chemical functionalization process. (d). Raman spectrum of a graphene device before (bottom curve) and after (top curve) functionalization.

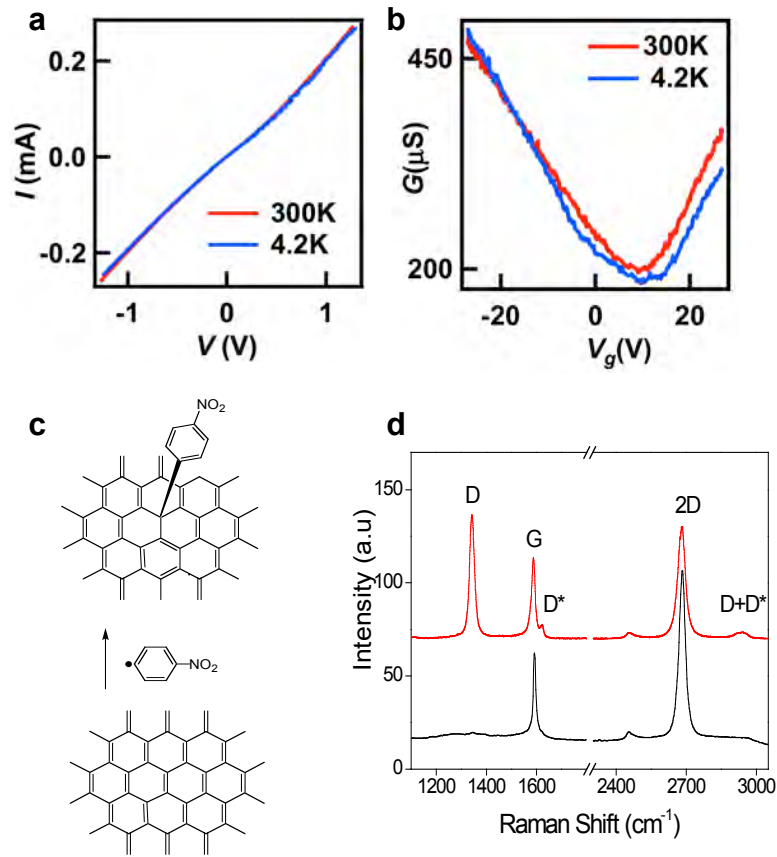

FIGURE 2. Transport data of NS1 after functionalization. (a,b). *I-V* and $G(V_g)$ characteristics. Red and blue traces are taken at 300K and 4K, respectively. (c). *dI/dV* (color) *vs.* bias *V* and $V_g$ at *T*=4K. (d). Line traces of (c) at $V_g$=38V (Dirac point) and -28.5V, respectively.

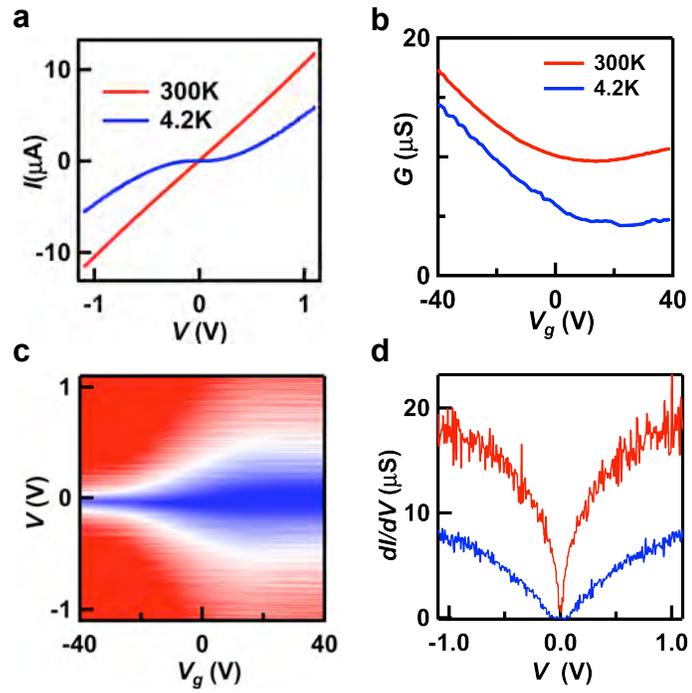

FIGURE 3. *T*-dependent transport data of NS1. (a). *I-V* curves of functionalized NS1 at *T*=14, 30, 60, 90, 120, 150, 178, 227, 257 and 283K. The data are taken at the Dirac point. (b). Linear response *G* vs. *1/T* for data taken at the Dirac point (red dots) and highly doped regimes (green dots). The solid lines are best-fitted curves to Eq. (1). Data at *T*=4K and 14K are not shown, since the measured $G<10^{-8}$ $\Omega^{-1}$ is limited by the instrument noise floor. (c). The same data points in (b), plotted as *G* vs. $T^{-1/2}$. The solid lines are curves fitted to Eq. (1) with $\alpha=1/2$. (d). *I* vs. reduced voltage $V-V_t$ for data taken at the Dirac point and *T*=4K and 14K, which collapse into a single curve. The red and green lines (indistinguishable) are best fits to Eq. (3), with exponent $\gamma=1.6$.

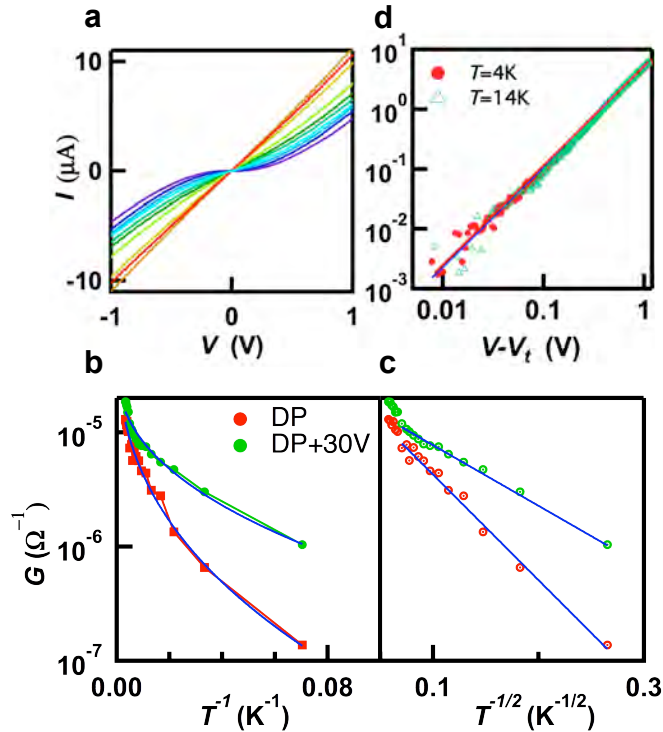

FIGURE 4. (a). False-color SEM image of a suspended graphene device. (b). $G(V_g)$ of a typical suspended device before functionalization. Scale bar: 1 μm. (c). I-V curves of a suspended functionalized device at $V_g=0$ and $T=300K$ (red curve, right axis) and 4K (blue curve, left axis), respectively. (d). Linear response $G$ vs. $1/T$. The solid line is the best-fit to $G(T)=G_0+A\exp(-E_A/k_BT)$, where $E_A \sim 40$ mV.

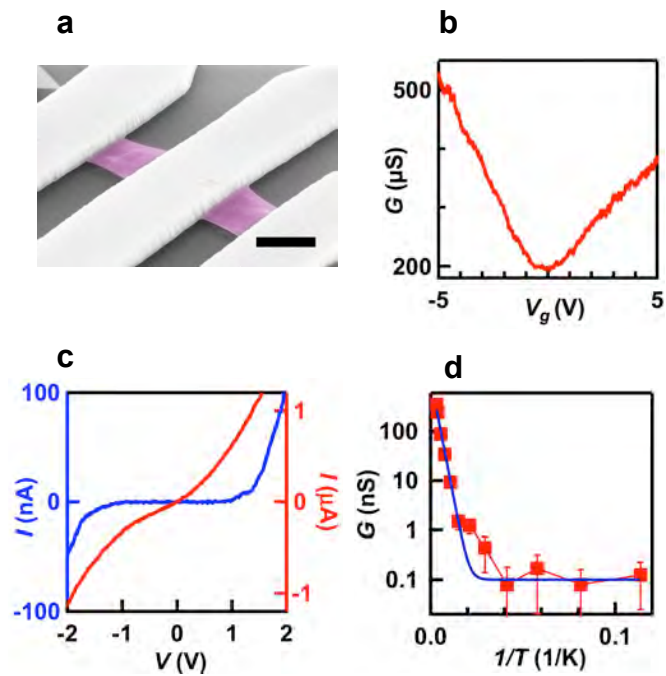